# Properties and superconductivity in Ti-doped NiTe$_2$ single crystals

B. S. de Lima,[a] R. R. de Cassia,[a] F. B. Santos,[a] L. E. Correa,[a] T. W. Grant,[a] A. L. R. Manesco,[a] G. W. Martins,[a] L. T. F. Eleno,[a] M. S. Torikachvili,[b] and A. J. S. Machado.[a]

(a) Escola de Engenharia de Lorena - Universidade de São Paulo, Lorena - SP, 12600-970, Brazil

(b) Department of Physics, San Diego State University, San Diego, California 92182-1233, USA

## ABSTRACT

Transition metal dichalcogenides (TMDs) usually show simple structures, however, with interesting properties. Recently some TMDs have been pointed out as type-II Dirac semimetals. In the present work, we investigate the physical properties of a new candidate for type-II Dirac semimetal and investigate the effect of titanium doping on physical properties of Ti-doped single crystalline samples of NiTe$_2$. It was found that this compound shows superconducting properties with a critical temperature close to 4.0 K. Interestingly, applied pressures up to 1.3 GPa have no effect upon the superconducting state. Density Functional Theory (DFT) calculations demonstrate the presence of a Dirac cone in the band structure of NiTe$_2$ literature when Spin-Orbit Coupling (SOC) is included, which is in agreement with a recent report for this compound. Also, our calculations demonstrate that Ti suppresses the formation of these non-trivial states.

I. INTRODUCTION

Transition metal dichalcogenides (TMDs) are compounds that crystallize in two-dimensional structures bonded together through Van der Waals forces between layers [1-5]. This leads to a great variety of physical properties, rich intercalation chemistry, and potential applications [6-8]. The low dimensional character of these TMds compounds often hosts electronic instabilities such as CDW transitions [9,10] or competition between CDW and superconducting ground states [11-13]. More recently, some TMDs have attracted attention because some of them host Dirac fermions and are classified as type-II Dirac or Weyl semimetals. For instance, $WTe_2$ was found to exhibit a large and non-saturating magneto-resistance [3] that was later pointed out as a consequence of band inversion and presence of Weyl nodes in the band structure [14,15]. $MoTe_2$ also was classified as a type-II Weyl semimetal [16] with a superconducting ground state at 0.1 K [17]. A recent ab-initio calculation article by B. Bradlyn *et al.* [18] suggests the existence of new 230 new compounds that should exhibit these non-trivial states, including $PtSe_2$, $PdSe_2$, $IrTe_2$, $HfTe_2$, and $NiTe_2$. Samples of $PdTe_2$ and $PtSe_2$ were already synthesized [19, 20] and had its electronic structured determined by ARPES which agreed with this study's predictions [18].

Within this context, this article is interested in the influence of Ti doping on the physical properties of nickel ditelluride, $NiTe_2$. This compound crystallizes in a 2D structural arrangement on the $CdI_2$ prototype structure, belonging to space group P-3m1 (164) with hexagonal symmetry [21]. Many TMDs compounds crystallize in this prototype, such as $HfTe_2$, $ZrTe_2$, $IrTe_2$, $CoTe_2$, $NiTe_2$, etc. In a previous article, we showed that, when copper is intercalated in the $ZrTe_2$ compound, in the nominal composition $Cu_{0.3}ZrTe_{1.2}$, superconductivity emerges from the intercalation [22]. In that article, it was shown that the superconducting transition is insensitivity to applied pressure and band structure calculations demonstrated some signatures of band inversion. In the present work, we present evidence that titanium intercalation in $NiTe_2$ induces superconductivity in Te deficient $NiTe_2$ samples. *Ab-initio* calculations demonstrate the presence of a Dirac cone in the band structure of which is in agreement with recent data published for this compound. Ti doping effects on the band structure are also discussed.

## II. EXPERIMENTAL PROCEDURE

Polycrystalline samples with $NiTe_2$, $NiTe_{1.5}$, and $Ti_{0.1}NiTe_{1.5}$ compositions were prepared from stoichiometric mixtures of high-purity Ni, Te, and Ti powders that were grounded thoroughly and pressed into pellets. These pellets were then encapsulated in a quartz tube under argon atmosphere, heated from room temperature to 750˚C and kept at this temperature for 48 hours. After this heat treatment, samples were reground, pressed into pellets, sealed in quartz tubes under the same conditions, and submitted to an additional heat treatment at the same temperature for 48 h. Finally, the tubes were quenched in iced water. The best polycrystalline samples, based upon X-ray diffraction data, were used in the crystal growth process. Single crystals were grown by melting the polycrystalline sample in quartz tubes that were placed in the furnace. These tubes containing the polycrystalline sample had the bottom in a conic shape. This side was then placed close to the furnace entrance and the other side of the tube close to the heating element in order to create a Bridgman-like gradient. The sample was maintained at 950˚C for 4 hours and then slowly cooled at 3˚C/min until 920˚C, then submitted to a new, slower cooling rate (2˚C/h) until 700˚C. When the sample reached 700˚C it was quenched in ice water. The single crystals produced had a metallic appearance and were easily cleaved. All the samples were characterized by x-ray diffraction (XRD) in a PANalytical diffractometer (model Empyrean), with a PIXcel3D detector using Cu-Kα radiation. X-ray diffraction data were analyzed using Rietveld refinement with the software PowderCell [23], Vesta Crystallography [24], and EXPGUI (GSAS) [25]. Magnetic and electric properties were measured using a Quantum Design Physical Property Measurement System (PPMS) EverCool II. The pressure dependence of the $Ti_{0.09}NiTe_{1.5}$ electrical resistivity at pressures up to 1.3 GPa was determined using a piston-cylinder self-locking cell, with silicone oil as the pressure-transmitting medium. The pressure at low temperatures was determined from the superconducting transition temperature of pure Pb.

## III. CALCULATION DETAILS

Density Functional Theory (DFT) [26] calculations of the ground state electronic structure were performed within the Kohn-Sham [27] scheme, using the scalar-relativistic Full-Potential-Linearized Augmented Plane Wave plus local orbitals (FP-LAPW+lo) method [28] as implemented in the exciting code [29]. Local Density Approximation [27] with the parameterization by Perdew and Wang [30] was chosen for the exchange-correlation functional, and spin-orbit coupling (SOC) was included. The lattice constants and atomic positions were relaxed until a convergence of $10^{-4}$ Ha on the total energy was reached. A mesh of 18x18x18 k-points in the Brillouin zone was adopted, with muffin-tin radii ($R_{MT}$) of 2.25 Bohr for all atoms and a product $R_{MT}K_{MAX}=10$ for the basis-size. The effects of Ti intercalation were simulated by considering the $TiNiTe_2$ (full Ti layer) stoichiometry.

## IV. RESULTS AND DISCUSSION

The binary Ni-Te phase diagram exhibits several intermediate phases. Among them is $NiTe_2$. It is stable with a large solubility range, between 52 and 67 at. % of tellurium [31]. This phase crystallizes with a $CdI_2$ prototype structure and is stable with a relatively large tellurium deficiency. In order to investigate this property and Ti doping two different samples were synthesized, with $NiTe_{1.5}$ and $Ti_{0.1}NiTe_{1.5}$ compositions. Figure 1 presents X-ray powder diffraction data for these samples.

The Rietveld refinement was carried out considering the space group P-3m1 (164), Ni and Te occupying 1a and 4d Wyckoff positions [21]. In Ti-doped samples, Ti was considered to occupy 1b Wyckoff positions (0, 0, ½). Adding Ti in this site does not create extra peaks in the original $NiTe_2$ ($CdI_2$ prototype) diffractogram. It is possible to observe an increase in the c parameter. This is experimental evidence that Ti was successfully intercalated into the Te-Te Van der Walls bonds. These considerations made the refinement converge fast and yield the error parameter goodness-of-fit ($\chi^2$) smaller than 2, which is an excellent value for X-ray diffraction analysis. Also, weighted-profile reliability $R_{WP}$ was found to be 8.68 and 11.59% in $NiTe_{1.5}$ and Ti-doped $NiTe_{1.5}$, respectively. The Ti intercalated polycrystalline samples with $\chi^2 \leq 2$ were used in the Bridgman-like method in order to obtain high-quality single crystals. X-ray diffraction data for a typical crystal grown by this method are shown in Figure 2.

Figure 2(a) shows that crystals cleave along the (00L) family of planes. This is expected since this is the direction related to the weak Van der Walls bonds. Figure 2(b) represents the rocking curve obtained from the (004) reflection and it shows a full width at half maximum (FWHM) of 0.09°, an extremely low value that vouches for the high quality of the crystal. In spite of the fact that there are no significant changes in the X-ray diffraction data of polycrystalline samples presented in Figure 1, the electrical behavior shows a strong dependence on the nominal composition of the sample. Figure 3 presents the electrical resistivity as a function of temperature for polycrystalline samples with a nominal composition of $NiTe_2$, $NiTe_{1.5}$ and a single crystal prepared from a polycrystalline sample of a nominal composition of $Ti_{0.1}NiTe_{1.5}$.

One can observe in Figure 3 that $NiTe_2$ sample (black spheres) displays a metal-like behavior. This behavior is in agreement with previous reports on band structure calculations due to several banding crossing the Fermi level [32]. For comparison, we measured also the resistivity of a sample with the composition of $NiTe_{1.5}$ (blue spheres). This sample also exhibits a metal-like behavior; however, one can observe an anomaly in the vicinity of 56 K. We have defined this temperature by the minimum point of the derivative dρ/dT vs T. The origin of this anomaly is not obvious and would require further investigation. However, it is possible that this anomaly may be related to CDW or even SDW ordering. These instabilities are relatively common in TMDs like NiTe2. Here we speculate that Te deficiency enhances the anomaly while Ti intercalation suppresses it at the same time as superconductivity emerges. This process occurs commonly in other TMDs upon chemical intercalation. See for instance $1T-Cu_XTiSe_2$ [7] and $6R-TaS_{2-x}Se_x$ [13]. Moreover, the Ti planes in $Ti_{0.1}NiTe_{1.5}$ have the effect of breaking the translational symmetry along the z-axis, a further indication that the 56K signal for $NiTe_{1.5}$ comes from CDW or SDW instability. The emergence of superconductivity also enforces that there is a competition for the ground state (i.e., at the Fermi level) between spontaneously broken translational and global phase invariance. It is clear that the anomaly around 56 K was suppressed and a superconducting state emerged at 4 K. Figure 4 presents further characterizations of this superconducting transition.

Figure 4(a) shows the magnetoresistance behavior with applied magnetic fields up to B = 0.7 T. The displacement observed in the behavior of the critical temperature with magnetic field proves the superconducting behavior observed at zero magnetic field. Here we argue that the resistance does not reach zero due to the non-homogeneity of Ti in the $NiTe_2$ matrix. For instance, a similar behavior has already been observed in other compounds that exhibit a superconductor state highly dependent on the dopant concentration, hence, the superconducting state will only happen in a well-defined Ti-concentration that can vary along the crystal [33]. Using the data from Figure 4(a), we can build the phase diagram and estimate the upper critical field of this compound. This diagram is shown in Figure 4(b). This diagram suggests that the upper critical field for this compound is close to $BC_2(0) \sim 0.9$ T, and one can estimate a Coherence Length of about 19.14 nm. This is relatively large when compared with others type-II superconducting compounds. Figure 4(c) presents measurements of the electrical resistivity as a function of temperature under pressure in the range 0–1.3 GPa.

In Figure 4(c) one can observe that, surprisingly, the superconducting transition temperature ($T_C$) is completely insensitive to applied pressure within this pressure range. This behavior also was observed for $Cu_{0.3}ZrTe_2$ [22], which crystallizes in the same prototype structure. In that case, band structure calculations strongly suggested signatures of topological effects represented by band touching or band inversion features in specific directions of the Brillouin zone when Spin-Orbit coupling is included. However, in the case of Cu doped $ZrTe_2$, these effects happen away from the Fermi level. Band structure calculations are shown for $NiTe_2$ and $TiNiTe_2$ in Figure 5.

Figure 5(a) shows the band structure calculation for $NiTe_2$ without titanium intercalation. These results reveal an electronic structure of a Dirac type II semimetal compound. In the direction D, this compound exhibit a Dirac cone-like feature, where the energy disperses linearly with k. Also, one can observe that the conduction and valence bands touch exactly at the Fermi level, which is a characteristic of Dirac fermions, similarly as in graphene. The apparent topological features shown in the band structure calculations suggest that some superconducting order parameter emerging in the system may lead to a topological superconducting phase. Thus, the independence of the superconductivity with the applied pressure may be a manifestation of the well-known robustness of symmetry protected topological phases against external perturbations such as pressure. This is consistent with recent theoretical predictions for NiTe2 [18], and analogous to results on other materials like PtSe$_2$ and PtTe$_2$ [19-20]. The main difference is that, in NiTe$_2$, the Dirac cone appears at the Fermi level, while on the Pt compounds it is found at approximately 1eV above the Fermi energy. When Titanium planes are intercalated between two van der Waals-bonded Te sheets, the electronic structure changes considerably. The conduction and valence bands touch each other near the Fermi level (with zero gaps) in between K-Γ and L-H directions, without considering the Spin-Orbit Coupling (SOC). However, when SOC is considered, these features are gapped in these directions, as shown in Figure 5(c), consequently, suppressing the Dirac-cone like feature present in undoped NiTe$_2$.

## V. CONCLUSIONS

This work presents a systematic study of the physical properties of Te deficient samples of the two-dimensional conductor NiTe$_2$. X-ray diffraction data shows that Ti can be intercalated in Te-Te Van der Waals gaps and consequently a superconductor state emerges in the vicinity of 4 K that is insensitive to pressure. Band structure calculations show nodule regions between the valence band and conduction band, where E disperses linearly with K for NiTe$_2$. This is what is expected for Dirac semimetals. Furthermore, by combining our experimental and calculations results we show that Ti intercalation suppresses the Dirac-cone like feature in the band structure and creates a superconductor ground state.

## ACKNOLOWDGMENTS

This work is based upon financial support by the Brazilian research agencies CAPES, CNPq (302850/2014-7, 443385/2014-9 and 142016/2013-6) and FAPESP (2013/16873-3, 2016/10167-8).

**FIGURE CAPTIONS**

**Figure 1** – X-ray diffraction data for samples of nominal composition NiTe1.5 and Ti0.1NiTe1.5. Structure representation is shown in the right side of the picture. Blue, yellow and purple spheres represent Ni, Te and Ti respectively.

**Figure 2** – (a) X-ray diffraction data obtained in a cleave surface of a single crystal. (b) Rocking curve obtained from the (004) reflection

**Figure 3** – Comparison between the electrical behaviors of samples with different compositions. Black and blue spheres represent the electrical dependence of polycrystals $NiTe_2$ and $NiTe_{1.5}$ respectively. Orange spheres are data obtained from a single crystal prepared after a sample of composition $Ti_{0.1}NiTe_{1.5}$. An anomaly can be observed in the Te deficient sample ($NiTe_{1.5}$). Ti doping suppresses this anomaly and induces a superconducting state to emerges at 4 K.

**Figure 4** – Superconducting properties of $Ti_{0.1}NiTe_{1.5}$ single crystal.

**Figure 5** – Calculated electronic band structures for (a) $NiTe_2$ without intercalation; (b) $TiNiTe_2$ without SOC; (c) $TiNiTe_2$ considering SOC.

**FIGURE_1**

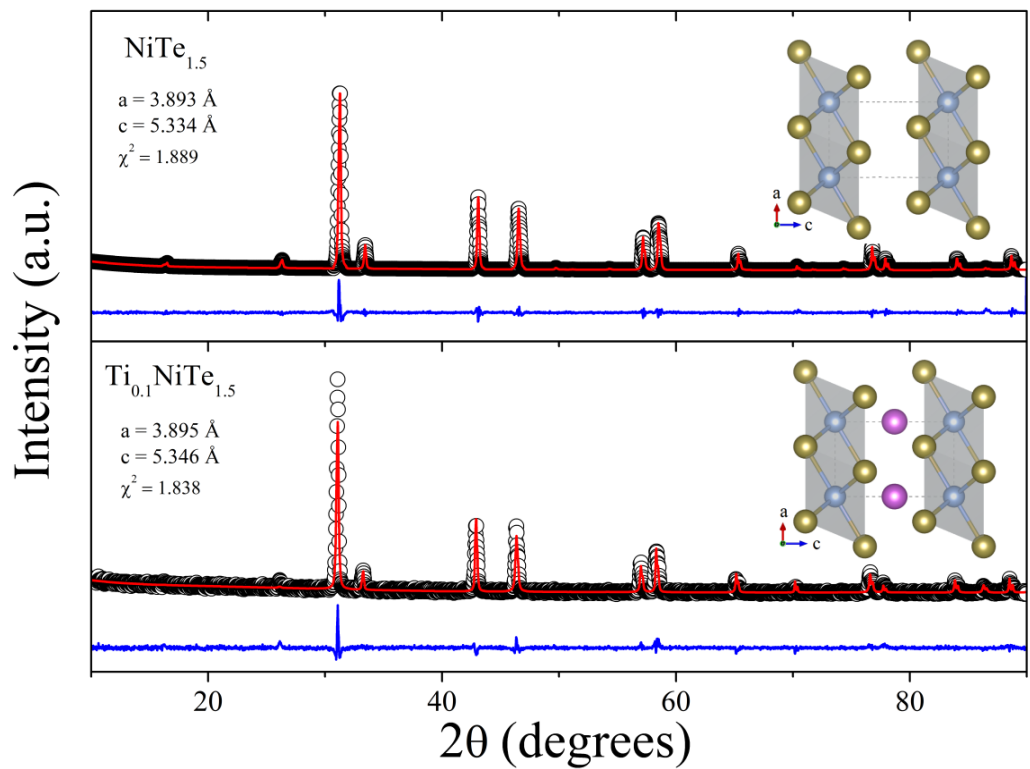

**FIGURE_2**

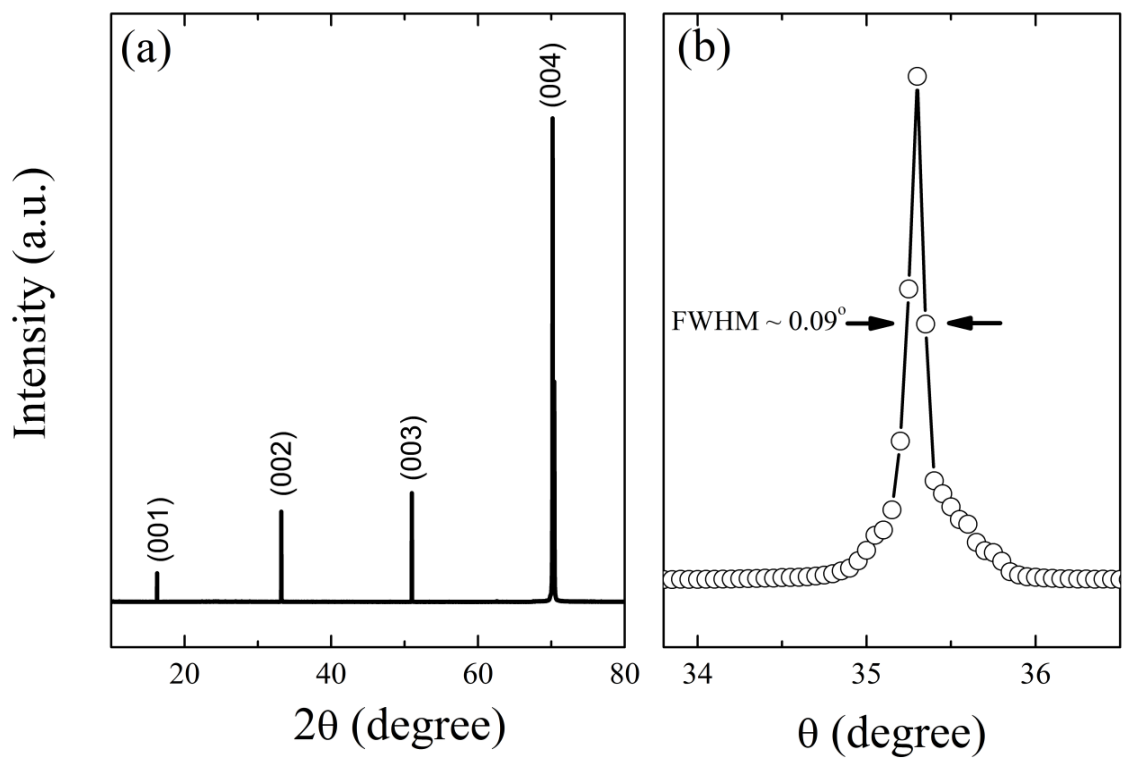

**FIGURE_3**

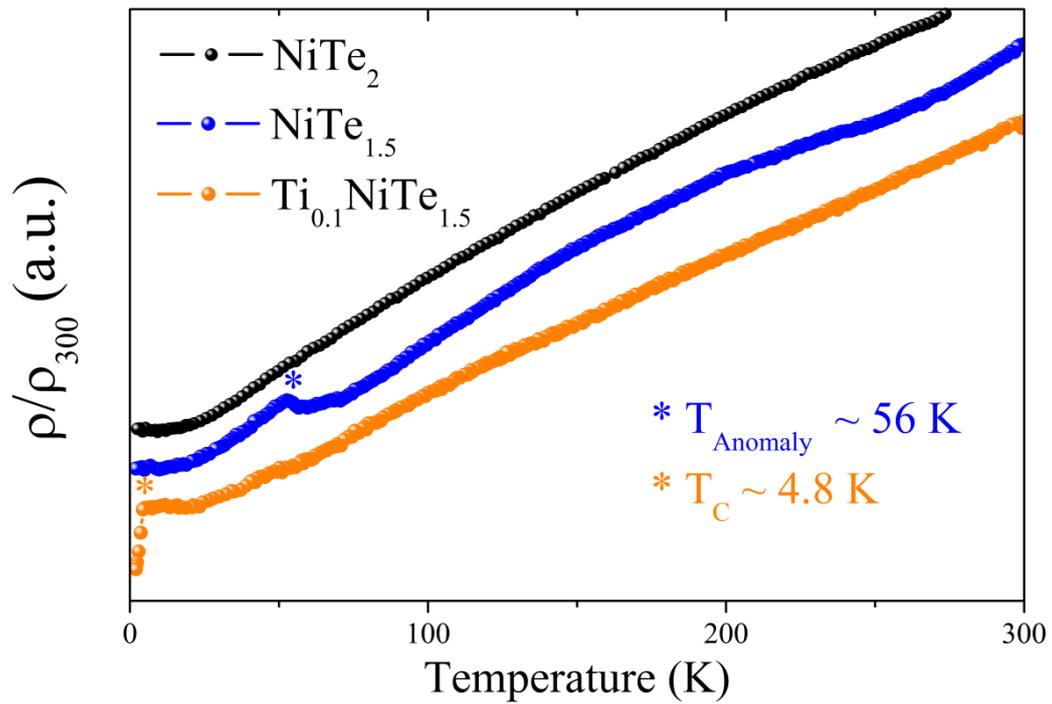

**FIGURE_4**

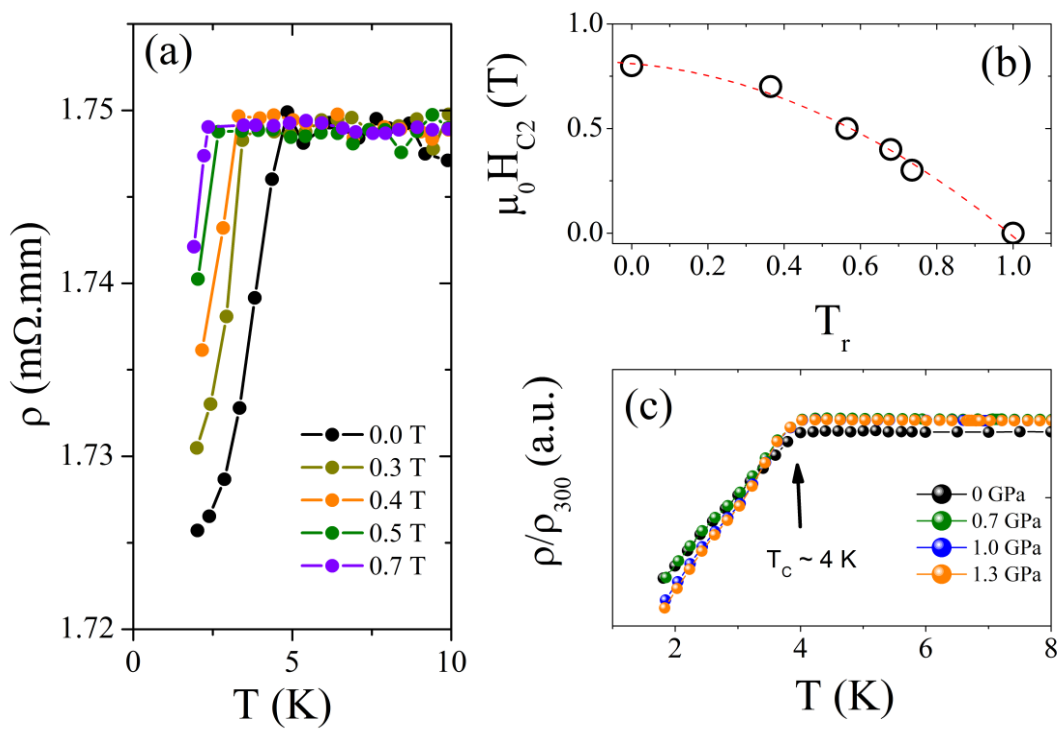

FIGURE_5

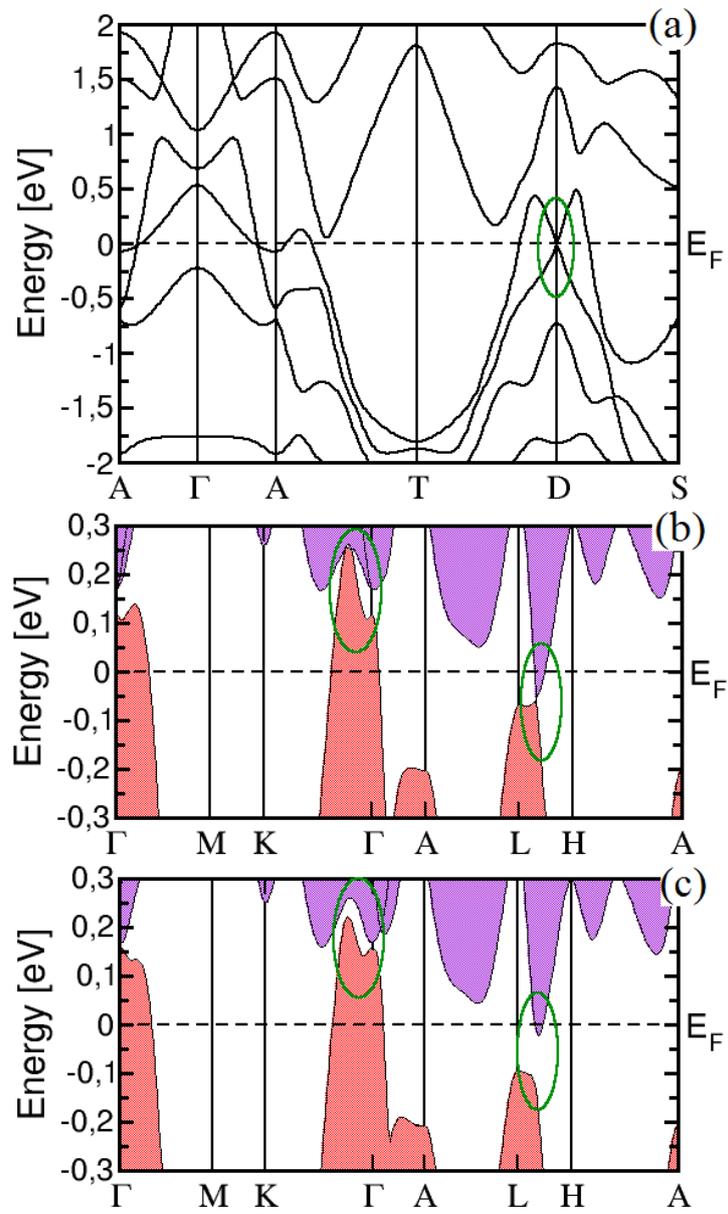